\documentclass[superscriptaddress,reprint,amsmath,amssymb,aps,floatfix]{revtex4-1}

\usepackage{amsmath,gensymb,textcomp,bm,dcolumn,eurosym,array,tabu,multirow,nicefrac,color,subfigure,graphicx,upgreek}
\usepackage[colorlinks, linkcolor=blue, citecolor=blue, urlcolor=blue, breaklinks=true]{hyperref}

\usepackage{mathpazo,times} 

\newcommand{\ket}[1]{\lvert #1 \rangle}

\usepackage[subsectionbib]{bibunits}
\usepackage{graphicx}
\usepackage{dcolumn}
\usepackage{bm,MnSymbol}
\usepackage{units}
\usepackage{psfrag}
\usepackage{float}
\usepackage{color}

\begin{document}
\title{Quantum interferometric two-photon excitation spectroscopy}

\author{Yuanyuan Chen}
\email{chenyy@xmu.edu.cn}
\affiliation{Department of Physics and Collaborative Innovation Center for Optoelectronic Semiconductors and Efficient Devices, Xiamen University, Xiamen 361005, China}
\author{Roberto de J. Le\'on-Montiel}
\email{roberto.leon@nucleares.unam.mx}
\affiliation{Instituto de Ciencias Nucleares, Universidad Nacional Aut\'onoma de M\'exico, Apartado Postal 70-543, 04510 Cd. Mx., M\'exico}
\author{Lixiang Chen}
\email{chenlx@xmu.edu.cn}
\affiliation{Department of Physics and Collaborative Innovation Center for Optoelectronic Semiconductors and Efficient Devices, Xiamen University, Xiamen 361005, China}

\begin{abstract}
Two-photon excitation spectroscopy is a nonlinear technique that has gained rapidly in interest and significance for studying the complex energy-level structure and transition probabilities of materials. While the conventional spectroscopy based on tunable classical light has been long established, quantum light provides an alternative way towards excitation spectroscopy with potential advantages in temporal and spectral resolution, as well as reduced photon fluxes. By using a quantum Fourier transform that connects the sum-frequency intensity and N00N-state temporal interference, we present an approach for quantum interferometric two-photon excitation spectroscopy. Our proposed protocol overcomes the difficulties of engineering two-photon joint spectral intensities and fine-tuned absorption-frequency selection. These results may significantly facilitate the use of quantum interferometric spectroscopy for extracting the information about the electronic structure of the two-photon excited-state manifold of atoms or molecules, in a ``single-shot'' measurement. This may be particularly relevant for photon-sensitive biological and chemical samples.
\end{abstract}
\maketitle

\section{Introduction}
Nonlinear spectroscopy techniques have been widely used in many disciplines, including but not limited to photoluminescence polymer and light-harvesting photosynthetic complexes \cite{Mukamel1995PrinciplesON,Hamm2011ConceptsAM,Hamm2011ConceptsAM}. Although these techniques are typically implemented by means of classical light---which is limited by shot-noise and may lack of robustness against experimental imperfections---recent investigations have suggested that quantum light may offer a new way to perform spectroscopy with provable advantages even in the presence of experimental non-idealities \cite{shi2020entanglement,schmidt2005Spectroscopy,Kira2011Quantum,Schlawin2021,cutipa2021}. Entanglement of photons is a prototypical example of such a quantum light, which lacks any counterpart in classical optics. Remarkably, the exploitation of quantum entanglement in two-photon absorption has discovered many fascinating phenomena, such as the linear dependence of two-photon absorption rate on the photon flux \cite{javanainen1990linear,landes2021experimental}, inducing disallowed atomic transitions \cite{muthukrishnan2004inducing}, manipulation of quantum pathways of matter \cite{roslyak2009multidimensional,schlawin2013Suppression}, as well as control in molecular processes \cite{shapiro2011generation}.

Among different quantum-enabled techniques, entangled two-photon absorption spectroscopy \cite{saleh,KOJIMA,nphoton,roberto_spectral_shape,dorfman2016,schlawin2017,oka2010,Schlawin-2018,villabona_calderon_2017,Varnavski2017,oka2018-1,oka2018-2,svozilik2018-1,svozilik2018-2,burdick2018,RobertoTemperatureControlled,Mukamel2020roadmap} has long been recognized as a promising spectroscopic tool for extracting the information about the electronic levels that participate in the two-photon excitation of a molecular samples. Although there is a current debate on the true quantum enhancement that such a technique might offer to spectroscopy \cite{raymer2021entangled, landes2021quantifying,raymer2021}, the experimental demonstration of its working principle, the so-called entangled two-photon absorption (eTPA), has become a topic of lively interest \cite{villabona2020, parzuchowski2021, tabakaev2021, samuel2021}. As one might expect, all this work has motivated the further development of novel spectroscopic tool proposals, particularly those that exploit the specificity of quantum light for resolving two-photon excited states \cite{paper2}.

Excitation spectroscopy is a powerful tool for extracting information about atomic or molecular excited states. It is typically implemented in the frequency domain by selecting a narrowband frequency slice of a broadband excitation light and recording the corresponding excited spectrum \cite{hipke2014broadband}. The requirement of scanning the excitation and/or emission wavelength generally leads to long acquisition times and complicated experimental setups. Interestingly, the simultaneous measurement of excitation spectra has proven to be beneficial for the study of energy transfer processes, as deviations between the two signals are indicative of loss channels. To tackle this, interferometric spectroscopy that can extract the excitation spectrum by performing a Fourier transform on its time-domain Mach-Zehnder or Michelson interference pattern has been well established \cite{Erling2019Single,piatkowski2016Broadband}. Nevertheless, the two-photon excitation spectroscopy remains relatively unexplored so far.

\begin{figure*}[t!]
\centering
\includegraphics[width=\linewidth]{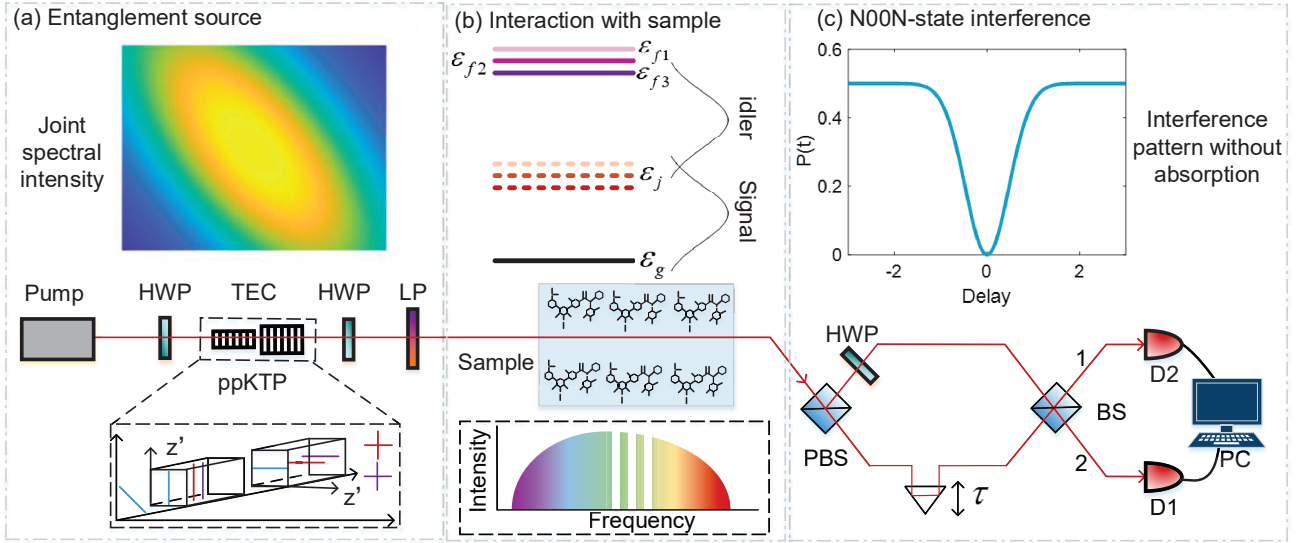}
\vspace{-10mm}
\caption{Experimental demonstration of quantum interferometric two-photon excitation spectroscopy. HWP, half-wave plate; TEC, temperature controller; ppKTP, periodically poled potassium titanyl phosphate crystal; LP, long pass filter; PBS, polarising beam splitter; BS: balanced beam splitter; sample: two-photon absorptive material, $D_{1/2}$, single photon detector. The inset in the entanglement source part illustrates the horizontally and vertically orientated crystals, which ensures that the incident diagonally-polarized  photons can pump these two crystals with equal probability. The inset in the interaction with sample part illustrates the two-photon absorption based on the incident photons with broadband sum-frequency. For example, three energy levels $\varepsilon_{f1-3}$ would absorb the corresponding two-photon components, and lead the residual photons to the interferometric measurement. The inset in the N00N-state interference part shows the original interference pattern without two-photon absorption.}
\label{figure_1}
\end{figure*}

Remarkably, quantum interferometry has been a long-standing setup for metrological and spectroscopic applications \cite{hong1987measurement,Mukamel2020roadmap,chou2020Frequency,chenglong2021}. Interestingly, N00N states, which refer to an extremal superposition of N quanta between two orthogonal modes as $(\ket{N,0}+\ket{0,N})/\sqrt{2}$, have been used as key resource for quantum enhanced metrology. In comparison to a single photon, these N-photon Fock states allow for an increased phase sensitivity scaling with photon number in longitudinal and rotational measurement \cite{hong2021quantum}. N00N-state interference reveals the fact that identical superposition modes $\ket{N,0}$ and $\ket{0,N}$ impinging on a balanced beam splitter from different input ports would bunch into a common output port or anti-bunch into two opposite output ports with corresponding probabilities that are determined by the relative phase factor between the two arms. Specifically, this sensitive phase can be introduced by optical delays between different paths of the interferometer \cite{lyons2018Attosecond,chen2019Hong}, or angular rotation between orbital angular momentum N00N states \cite{hiekkamki2021Photonic}, as well as frequency shift between N-photon sum-frequency \cite{roger2016coherent}. Thus, N00N-state interferometry invoking photonic degrees of freedom may pave the way towards unconditional super-sensitivity.

In this work, we express the N00N-state interference patterns in terms of biphoton sum-frequency intensities by using a quantum Fourier transform that provides a translation between spectral and temporal domains of the biphoton wavefunction. Since the sum-frequency of two photons, namely the excitation energy, determines the oscillation period of the N00N-state interference pattern, the temporal signals recorded at the output of the N00N-state interferometer inversely reveals its spectroscopic information. To certify its viability, we also present an experimental scheme that implements the quantum interferometric measurement of two-photon excitation spectrum, which provides a useful tool for quantum interferometric nonlinear spectroscopy.

In comparison to the conventional emission spectroscopy protocol, our Fourier approach allows for a rapid, and simultaneous, extraction of the sample's excitation spectrum, leading to a time-efficient two-photon excitation spectroscopy. This can be attributed to the exploitation of broadband pump laser without any requirement for precise and complicated scanning in the spectral domain. Additionally, the resultant intensity of emission photons may reduce the detection limit in the context of nonlinear spectroscopy. Instead, our approach measures the residual incident photons after two-photon absorption process, whose intensity remains relatively steady. Remarkably, we demonstrate that the sensitivity of this broadband technique is comparable to that of emission spectroscopy under a premise that the sufficient number of temporal signals in N00N-state interference is measured. As a direct result, our approach may inspire more practical applications in quantum spectroscopy and quantum metrology, in particular for photon-sensitive biological and chemical samples.

\section{Quantum interferometric spectroscopy with N00N states}
Our approach makes use of an entangled-photon source as the incident two-photon signal, and a N00N-state spatial interferometer to extract the specific absorption spectrum. We present a feasible experimental setup to fulfill the task of two-photon excitation as shown in Fig.\ \ref{figure_1}. The entangled photons are generated by pumping the nonlinear crystals with a broadband laser. In this spontaneous parametric down conversion (SPDC) process, a strong pump photon would spontaneously decay into a pair of daughter photons, known as signal and idler photons with the energy conservation as $\omega_s+\omega_i=\omega_p$, where $\omega_s$, $\omega_i$, and $\omega_p$ are the frequencies of signal, idler and pump, respectively. As a direct result of the broadband laser, the joint spectral intensity shows that the down-converted photons are not strictly frequency correlated [see the inset in Fig.\ \ref{figure_1}\textcolor{blue}{(a)}]. A pair of nonlinear crystals (e.g., ppKTP) are placed in sequence, whereby the optical axis of the second crystal is rotated by $90^\degree$ with respect to the first crystal \cite{chen2018polarization}. Both of these two nonlinear crystals are designed for identical collinear phase matching, such that the photons originating from two crystals are indistinguishable. Note that for the sake of simplicity, and to accomplish indistinguishability, we consider a type-0 phase matching. Balanced pumping enables equal probability amplitudes from SPDC emission $\ket{H}\rightarrow\ket{HH}$ in the first crystal, and $\ket{V}\rightarrow\ket{VV}$ in the second crystal, where $\ket{H}$ and $\ket{V}$ represent horizontal and vertical polarization, respectively. The coherent superposition of pair generation possibilities in two crystals results in a polarization entanglement as
\begin{equation}
\ket{\Psi}=(\ket{HH}+\ket{VV})/\sqrt{2}.
\end{equation}
Experimentally, a long pass filter is further needed in order to eliminate the residual pump light. Then, the down-converted photons are sent to interact with the target sample.

\begin{figure}[!t]
\centering
\subfigure[]{
\label{Fig2.sub.1}
\includegraphics[width=\linewidth]{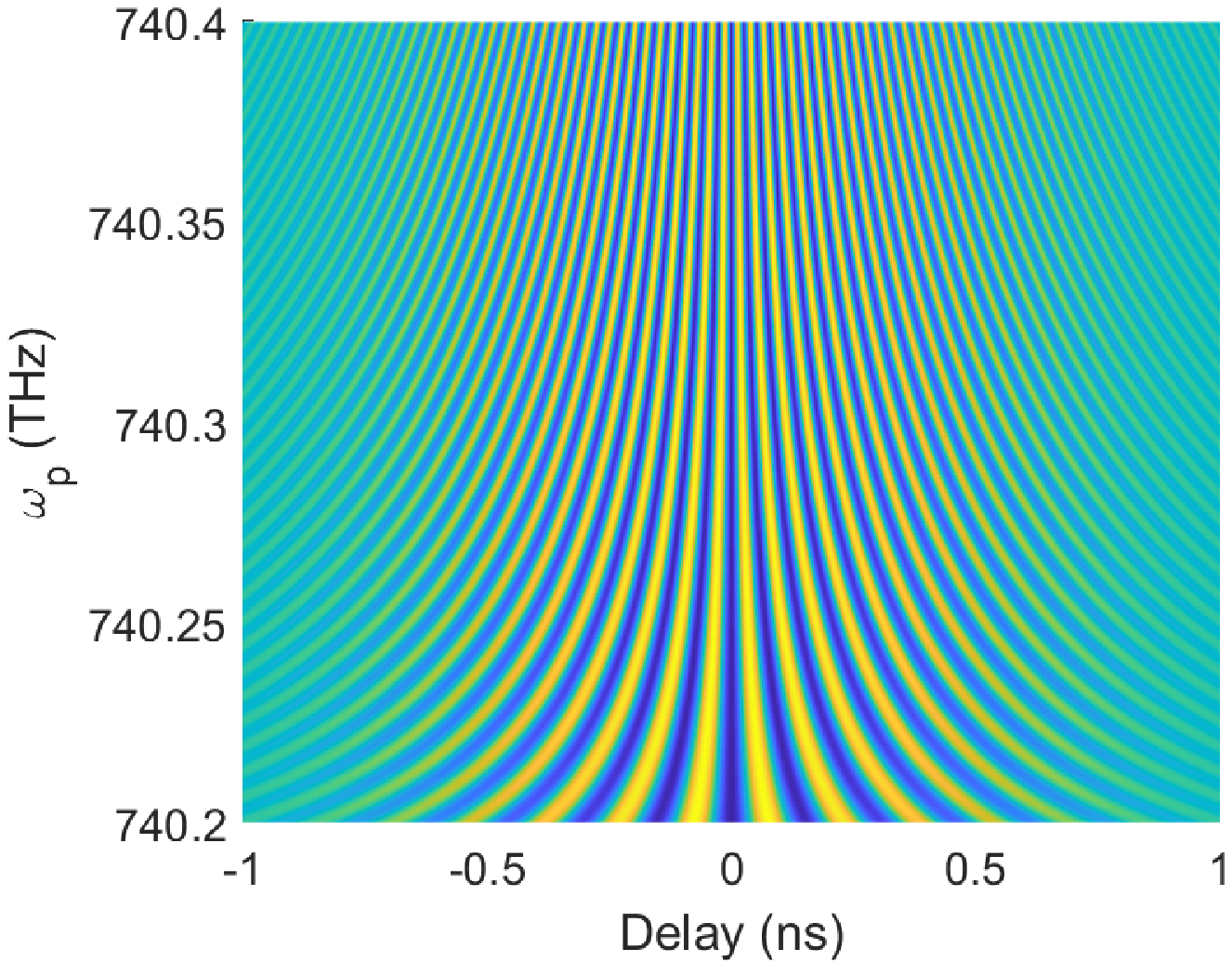}}
\subfigure[]{
\label{Fig2.sub.2}
\includegraphics[width=\linewidth]{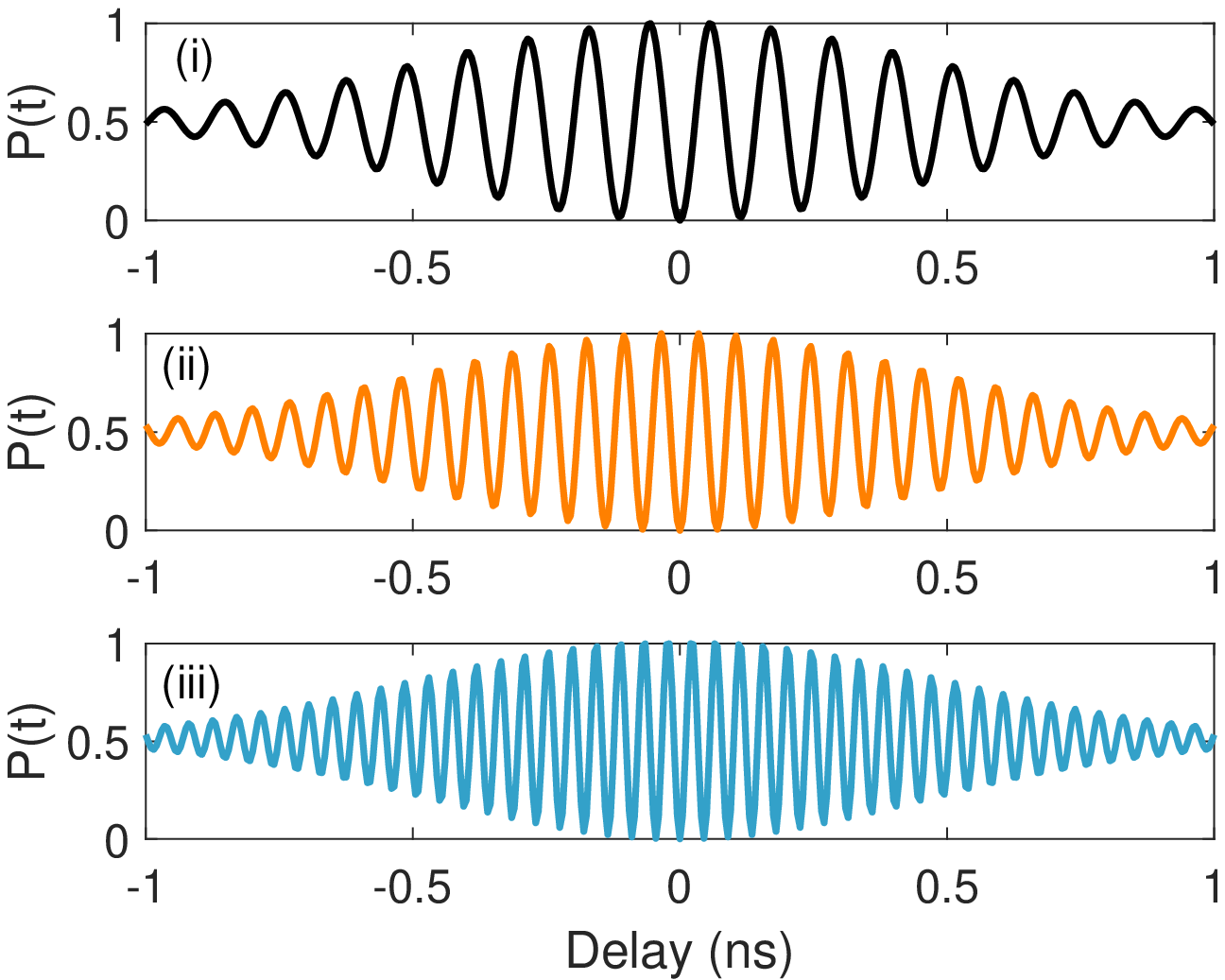}}
\caption{(a) The normalized coincidence counts versus sum-frequency of two-photon and relative time delay introduced in the unbalanced interferometer. (b) The N00N-state interference pattern for specific sum-frequencies (i) $\omega_p$=\unit[740.215]{THz}, (ii) $\omega_p$=\unit[740.250]{THz}, and (iii) $\omega_p$=\unit[740.300]{THz}, which reveals that the sum-frequency determines the oscillation period within a typically Gaussian envelope.}
\label{figure_2}
\end{figure}

In the interaction process, we assume that the medium is initially in its ground state $\ket{g}$ (with energy $\varepsilon_g$). The incident two-photon components drive an excitation of the medium to the final state $\ket{f}$ (with energy $\varepsilon_f$). Note that second-order time-dependent perturbation theory predicts that two-photon excitation may proceed through the intermediate state $\ket{j}$ (with energy $\varepsilon_j$) \cite{Mukamel1995PrinciplesON}. By using such theory, one can find that the two-photon absorption signal is expressed as \cite{RobertoTemperatureControlled}
\begin{equation}
\mathcal{P}_{g\rightarrow f}(\omega_{p_j})=\sum_{j}|\delta(\frac{\omega_{p_j}-\varepsilon_f}{4\pi})|^2F_{p_j},
\end{equation}
where $F_{p_j}$ is the absorption factor for individual pump frequency $\omega_{p_j}$. As shown in Fig.\ \ref{figure_1}\textcolor{blue}{(b)}, the two-photon component with specific sum-frequency that equals to $\varepsilon_f$ are absorbed, while the residual two-photon components are transmitted through this transparent sample.

After the quantum light-matter interaction, a polarizing beam splitter (PBS) is used to deterministically route paired polarization-entangled photons into two distinct spatial modes. Thus, the component in superposition state $\ket{HH}$ is transmitted through the PBS, and the component in superposition state $\ket{VV}$ is reflected by PBS and pass through a translation stage, which can be used to scan its arriving time. To enable interaction between photon pairs, a half-wave plate (HWP) is placed in one of the output ports of the PBS such that the distinguishability in polarization domain is eliminated. Thus, the paired photons impinge on a balanced beam splitter from opposite input modes, which constitutes a N00N-state interferometer. The non-classical N00N-state interference pattern can then be recorded by identifying coincidence counts at distinct output ports. Specifically, the down-converted photons are detected by silicon avalanche photodiodes, and two-fold events are identified using a fast electronic AND gate when two photons arrive at the detectors within a coincidence window. According to the calculation, the N00N-state interference can be written as \cite{jin2018Extended}
\begin{equation}\label{eq: interference}
\begin{split}
P(t)=&\frac{1}{2}[1+\int_{-\infty}^{\infty}\int_{-\infty}^{\infty}d\omega_sd\omega_i\\
&|f(\omega_s,\omega_i)|^2\exp[-i(\omega_s+\omega_i)t]],
\end{split}
\end{equation}
where $f(\omega_s,\omega_i)$ is a joint spectral intensity for the signal photon with a frequency of $\omega_s$ and idler photon with a frequency of $\omega_i$. This equation reveals that the oscillation period of the N00N-state interference pattern is directly determined by $\omega_s+\omega_i=\omega_p$, and not on $\omega_s$ or $\omega_i$, individually (see Fig. \ref{figure_2}). In other words, the two-photon absorption spectrum, resulting from the quantum light-matter interaction [see Fig.\ \ref{figure_1}\textcolor{blue}{(b)}], would make contributions to the N00N-state interference pattern.

\begin{figure*}[t!]
\centering
\includegraphics[width=\linewidth]{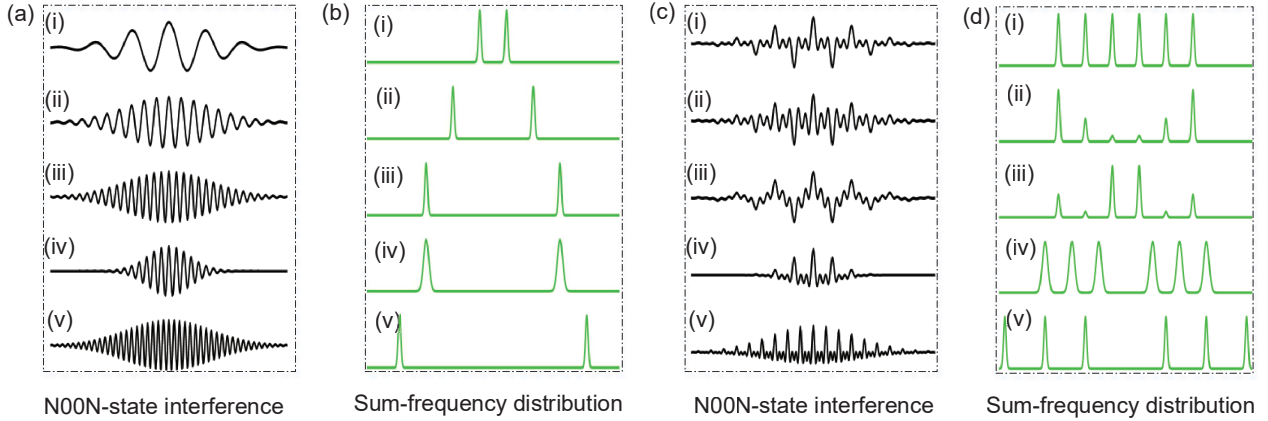}
\caption{In the quantum interferometric two-photon excitation spectroscopy, (a) the input signal is defined as N00N-state interference patterns. By performing a Fourier transform on their second-order correlation functions, (b) the calculated spectral signal reveals the oscillation period, namely the separation distance between two symmetric peaks is proportional to the sum-frequency intensity of paired photons. (c) If the input is a superposition of multiple simple sinusoidal curves, the Fourier transform on their second-order correlation functions reveals the (d) spectral signal that the incident sum-frequency intensity carries.}
\label{figure_3}
\end{figure*}
Inversely, in order to obtain the absorption spectrum, we can decompose the temporal interference pattern in terms of two-photon sum-frequency intensity. As shown in Fig.\ \ref{figure_3}\textcolor{blue}{(a)}, let us now consider the measured N00N-state interference pattern, described by Eq. \eqref{eq: interference}, which manifests itself as a periodic oscillation within a typically Gaussian envelope. The second-order correlation function $G(t)$ is defined as
\begin{equation}\label{second-order correlation}
\begin{split}
G(t)=&1-2P(t)\\
=&\int_{-\infty}^{\infty}\int_{-\infty}^{\infty}d\omega_sd\omega_i|f(\omega_s,\omega_i)|^2\exp[-i(\omega_s+\omega_i)t]],
\end{split}
\end{equation}
which can be considered as a two-photon temporal signal. According to the energy conservation law $\omega_p=\omega_s+\omega_i$, Eq. \eqref{second-order correlation} is simplified to
\begin{equation}
G(t)=\int_{-\infty}^{\infty}d\omega_pF(\omega_p)\exp(-i\omega_pt),
\end{equation}
where $F(\omega_p)$ is the Gaussian spectral amplitude function that fulfills the normalization condition: $\int d\omega_pF(\omega_p)=1$. It is straightforward to realize that the second-order correlation function and sum-frequency spectrum intensity can be expressed in the form of a Fourier transform as
\begin{equation} \label{eq: spectrum}
\begin{split}
F(\omega_p)=\frac{1}{2\pi}\int_{-\infty}^\infty dt G_2(t) \exp(i\omega_pt).
\end{split}
\end{equation}
Since the two-photon temporal signal exhibits as a cosine oscillation, its corresponding sum-frequency intensity obtained by performing a Fourier transform is shown in Fig.\ \ref{figure_3}\textcolor{blue}{(b)}. These recovered spectral patterns manifest as two symmetric peaks, wherein the separation distance is determined by the oscillation period that equals to the pump frequency line [see Fig.\ \ref{figure_3}\textcolor{blue}{(a-b i,ii,iii)}]. In addition, the base-to-base envelope width in Fig.\ \ref{figure_3}\textcolor{blue}{(b)} reveals the coherence length, and it is inversely proportional to the pump frequency linewidth [see Fig.\ \ref{figure_3}\textcolor{blue}{(a-b iv,v)}].

Apart from the pump light that is characterized by a single linewidth frequency, broadband frequency combs have been used in a large number of applications such as two-photon frequency-comb spectroscopy \cite{grinin2020Two,ozawa2012chirped}. If the pump light contains multiple frequency lines, the measured N00N-state interference pattern would be a consequence of the superposition of multiple interference patterns obtained from single pump frequency lines [see Fig.\ \ref{figure_3}\textcolor{blue}{(c)}]. Analogously, by performing a Fourier transform on its second-order correlation function, we can also recover the two-photon spectral distribution from its N00N-state interference pattern as shown in Fig.\ \ref{figure_3}\textcolor{blue}{(d)}, which is directly related to the two-photon absorption spectrum. In particular, we note that the weights of individual interference patterns are directly determined by its corresponding probability amplitude of pump frequency lines. These results show that our approach is also appropriate for complex pump modulation, which provides an alternative route towards broadband absorption spectroscopy.

\section{Quantum interferometric measurement of two-photon excitation spectra}
\begin{figure*}[!t]
\centering
\includegraphics[width=\linewidth]{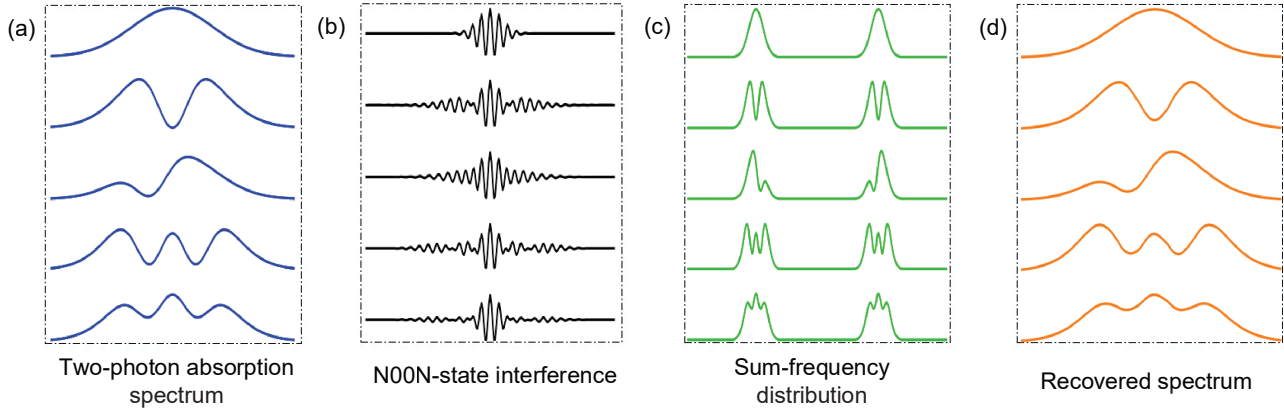}
\caption{(a) Two-photon absorption spectrum with complex structures. (b) The theoretical simulation of their N00N-state interference patterns. Analogously, by performing a Fourier transform on their second-order correlation functions, (c) the calculated spectral distribution can be used to recover the (d) original two-photon absorption spectrum.}
\label{figure_4}
\end{figure*}
Let us now consider a practical setting in the application of two-photon excitation spectroscopy. Instead of scanning the incident pump frequency lines, we use a pump light with broad bandwidth such that the two-photon absorption can be achieved without the requirement of precise scanning and complicated experimental setups (see the inset in Fig.\ \ref{figure_1}\textcolor{blue}{(b)}). The core of our method is to identify the measured spectrum as the one obtained by subtracting the two-photon absorption signal from the initially broad, incident pump. This pump light with broad spectral distribution is composed with a large number of linewidth pump. By making a Fourier transform on the second-order correlation function of the N00N-state interference signals, the recovered spectrum agrees well with the two-photon absorption spectrum as shown in Fig.\ \ref{figure_4}. Thereinto, the intensity of two-photon absorption can also be recovered since the Fourier transform enables us to extract the probability amplitude.

Interestingly, in a practical application, it is not generally allowed to measure two-photon excitation spectra with extremely narrow spectral linewidth. As shown in Eq. \eqref{eq: spectrum}, the resolution of our method is proportional to the measurement precision of temporal signals, namely the sampling spacing in temporal domain. Additionally, given that the Quantum Cram\'{e}r-Rao bound theory states that the standard deviation is inversely proportional to the number of independent trials of the experiment, the resolution can be further enhanced by repeating experimental measurements, increasing the brightness of entangled-photon source and detection efficiency, as well as decreasing the single-photon loss rate in transmission.

\section{Discussion}
We have presented a feasible scheme to implement quantum interferometric two-photon excitation spectroscopy. By performing a quantum Fourier transform on the second-order correlation function of the N00N-state interference pattern, one can extract the two-photon excitation spectrum of the sample without the necessity of engineering the two-photon joint spectral intensity. In addition, the use of a quantum source may provide additional advantages in temporal and spectral resolution, which can facilitate its use in hyperfine structures \cite{schlawin2013Suppression}. By combining our work with two-photon virtual state spectroscopy \cite{RobertoTemperatureControlled}, the complete diagram of two-photon absorption could readily be obtained.

The most remarkable outcome of this study is the fact that we provide a connection between sum-frequency intensity and N00N-state interference pattern. In addition to quantum interferometric spectroscopy, which extracts spectral signals from temporal measurements, spectral-domain optical coherence tomography that detects the thickness induced by a transparent material may also be accessible \cite{marchand2021soliton,pablo2020spectrally}. This might motivate the development of novel applications of full-field optical coherence tomography and three-dimensional imaging. In conclusion, we believe that fully harnessing quantum interferometric spectroscopy provides additional tools that ultimately may broaden the path towards practical quantum information processing and quantum metrology.

\begin{acknowledgments}
This work is supported by the National Natural Science Foundation of China (NSFC) (12034016, 12004318, 61975169), the Fundamental Research Funds for the Central Universities at Xiamen University (20720190057), the Natural Science Foundation of Fujian Province of China (2020J05004), the Natural Science Foundation of Fujian Province of China for Distinguished Young Scientists (2015J06002), and the program for New Century Excellent Talents in University of China (NCET-13-0495). R.J.L.-M. thankfully acknowledges financial support by CONACyT under the project CB-2016-01/284372 and by DGAPA-UNAM under the project UNAM-PAPIIT IN102920.
\end{acknowledgments}

\bibliography{biblio}

\begin{thebibliography}{55}%
\makeatletter
\providecommand \@ifxundefined [1]{%
 \@ifx{#1\undefined}
}%
\providecommand \@ifnum [1]{%
 \ifnum #1\expandafter \@firstoftwo
 \else \expandafter \@secondoftwo
 \fi
}%
\providecommand \@ifx [1]{%
 \ifx #1\expandafter \@firstoftwo
 \else \expandafter \@secondoftwo
 \fi
}%
\providecommand \natexlab [1]{#1}%
\providecommand \enquote  [1]{``#1''}%
\providecommand \bibnamefont  [1]{#1}%
\providecommand \bibfnamefont [1]{#1}%
\providecommand \citenamefont [1]{#1}%
\providecommand \href@noop [0]{\@secondoftwo}%
\providecommand \href [0]{\begingroup \@sanitize@url \@href}%
\providecommand \@href[1]{\@@startlink{#1}\@@href}%
\providecommand \@@href[1]{\endgroup#1\@@endlink}%
\providecommand \@sanitize@url [0]{\catcode `\\12\catcode `\$12\catcode
  `\&12\catcode `\#12\catcode `\^12\catcode `\_12\catcode `\%12\relax}%
\providecommand \@@startlink[1]{}%
\providecommand \@@endlink[0]{}%
\providecommand \url  [0]{\begingroup\@sanitize@url \@url }%
\providecommand \@url [1]{\endgroup\@href {#1}{\urlprefix }}%
\providecommand \urlprefix  [0]{URL }%
\providecommand \Eprint [0]{\href }%
\providecommand \doibase [0]{http://dx.doi.org/}%
\providecommand \selectlanguage [0]{\@gobble}%
\providecommand \bibinfo  [0]{\@secondoftwo}%
\providecommand \bibfield  [0]{\@secondoftwo}%
\providecommand \translation [1]{[#1]}%
\providecommand \BibitemOpen [0]{}%
\providecommand \bibitemStop [0]{}%
\providecommand \bibitemNoStop [0]{.\EOS\space}%
\providecommand \EOS [0]{\spacefactor3000\relax}%
\providecommand \BibitemShut  [1]{\csname bibitem#1\endcsname}%
\let\auto@bib@innerbib\@empty
\bibitem [{\citenamefont {Mukamel}(1995)}]{Mukamel1995PrinciplesON}%
  \BibitemOpen
  \bibfield  {author} {\bibinfo {author} {\bibfnamefont {S.}~\bibnamefont
  {Mukamel}},\ }\href@noop {} {\emph {\bibinfo {title} {Principles of Nonlinear
  Optical Spectroscopy}}}\ (\bibinfo {year} {1995})\BibitemShut {NoStop}%
\bibitem [{\citenamefont {Hamm}\ and\ \citenamefont
  {Zanni}(2011)}]{Hamm2011ConceptsAM}%
  \BibitemOpen
  \bibfield  {author} {\bibinfo {author} {\bibfnamefont {P.}~\bibnamefont
  {Hamm}}\ and\ \bibinfo {author} {\bibfnamefont {M.}~\bibnamefont {Zanni}},\
  }\href@noop {} {\emph {\bibinfo {title} {Concepts and Methods of 2D Infrared
  Spectroscopy}}}\ (\bibinfo {year} {2011})\BibitemShut {NoStop}%
\bibitem [{\citenamefont {Shi}\ \emph {et~al.}(2020)\citenamefont {Shi},
  \citenamefont {Zhang}, \citenamefont {Pirandola},\ and\ \citenamefont
  {Zhuang}}]{shi2020entanglement}%
  \BibitemOpen
  \bibfield  {author} {\bibinfo {author} {\bibfnamefont {H.}~\bibnamefont
  {Shi}}, \bibinfo {author} {\bibfnamefont {Z.}~\bibnamefont {Zhang}}, \bibinfo
  {author} {\bibfnamefont {S.}~\bibnamefont {Pirandola}}, \ and\ \bibinfo
  {author} {\bibfnamefont {Q.}~\bibnamefont {Zhuang}},\ }\href {\doibase
  10.1103/PhysRevLett.125.180502} {\bibfield  {journal} {\bibinfo  {journal}
  {Phys. Rev. Lett.}\ }\textbf {\bibinfo {volume} {125}},\ \bibinfo {pages}
  {180502} (\bibinfo {year} {2020})}\BibitemShut {NoStop}%
\bibitem [{\citenamefont {Schmidt}\ \emph {et~al.}(2005)\citenamefont
  {Schmidt}, \citenamefont {Rosenband}, \citenamefont {Langer}, \citenamefont
  {Itano}, \citenamefont {Bergquist},\ and\ \citenamefont
  {Wineland}}]{schmidt2005Spectroscopy}%
  \BibitemOpen
  \bibfield  {author} {\bibinfo {author} {\bibfnamefont {P.}~\bibnamefont
  {Schmidt}}, \bibinfo {author} {\bibfnamefont {T.}~\bibnamefont {Rosenband}},
  \bibinfo {author} {\bibfnamefont {C.}~\bibnamefont {Langer}}, \bibinfo
  {author} {\bibfnamefont {W.}~\bibnamefont {Itano}}, \bibinfo {author}
  {\bibfnamefont {J.}~\bibnamefont {Bergquist}}, \ and\ \bibinfo {author}
  {\bibfnamefont {D.}~\bibnamefont {Wineland}},\ }\href@noop {} {\bibfield
  {journal} {\bibinfo  {journal} {Science}\ }\textbf {\bibinfo {volume}
  {309}},\ \bibinfo {pages} {p. 749} (\bibinfo {year} {2005})}\BibitemShut
  {NoStop}%
\bibitem [{\citenamefont {Kira}\ \emph {et~al.}(2011)\citenamefont {Kira},
  \citenamefont {Koch}, \citenamefont {Smith}, \citenamefont {Hunter},\ and\
  \citenamefont {Cundiff}}]{Kira2011Quantum}%
  \BibitemOpen
  \bibfield  {author} {\bibinfo {author} {\bibfnamefont {M.}~\bibnamefont
  {Kira}}, \bibinfo {author} {\bibfnamefont {S.~W.}\ \bibnamefont {Koch}},
  \bibinfo {author} {\bibfnamefont {R.~P.}\ \bibnamefont {Smith}}, \bibinfo
  {author} {\bibfnamefont {A.}~\bibnamefont {Hunter}}, \ and\ \bibinfo {author}
  {\bibfnamefont {S.~T.}\ \bibnamefont {Cundiff}},\ }\href@noop {} {\bibfield
  {journal} {\bibinfo  {journal} {Nature Physics}\ }\textbf {\bibinfo {volume}
  {7}},\ \bibinfo {pages} {800} (\bibinfo {year} {2011})}\BibitemShut {NoStop}%
\bibitem [{\citenamefont {S\'anchez Mu\~noz}\ \emph {et~al.}(2021)\citenamefont
  {S\'anchez Mu\~noz}, \citenamefont {Frascella},\ and\ \citenamefont
  {Schlawin}}]{Schlawin2021}%
  \BibitemOpen
  \bibfield  {author} {\bibinfo {author} {\bibfnamefont {C.}~\bibnamefont
  {S\'anchez Mu\~noz}}, \bibinfo {author} {\bibfnamefont {G.}~\bibnamefont
  {Frascella}}, \ and\ \bibinfo {author} {\bibfnamefont {F.}~\bibnamefont
  {Schlawin}},\ }\href {\doibase 10.1103/PhysRevResearch.3.033250} {\bibfield
  {journal} {\bibinfo  {journal} {Phys. Rev. Research}\ }\textbf {\bibinfo
  {volume} {3}},\ \bibinfo {pages} {033250} (\bibinfo {year}
  {2021})}\BibitemShut {NoStop}%
\bibitem [{\citenamefont {Cutipa}\ and\ \citenamefont
  {Chekhova}(2021)}]{cutipa2021}%
  \BibitemOpen
  \bibfield  {author} {\bibinfo {author} {\bibfnamefont {P.}~\bibnamefont
  {Cutipa}}\ and\ \bibinfo {author} {\bibfnamefont {M.~V.}\ \bibnamefont
  {Chekhova}},\ }\href@noop {} {\enquote {\bibinfo {title} {Bright squeezed
  vacuum for two-photon spectroscopy: simultaneously high resolution in time
  and frequency, space and wavevector},}\ } (\bibinfo {year} {2021}),\ \Eprint
  {http://arxiv.org/abs/2110.12832} {arXiv:2110.12832 [quant-ph]} \BibitemShut
  {NoStop}%
\bibitem [{\citenamefont {Javanainen}\ and\ \citenamefont
  {Gould}(1990)}]{javanainen1990linear}%
  \BibitemOpen
  \bibfield  {author} {\bibinfo {author} {\bibfnamefont {J.}~\bibnamefont
  {Javanainen}}\ and\ \bibinfo {author} {\bibfnamefont {P.~L.}\ \bibnamefont
  {Gould}},\ }\href {\doibase 10.1103/PhysRevA.41.5088} {\bibfield  {journal}
  {\bibinfo  {journal} {Phys. Rev. A}\ }\textbf {\bibinfo {volume} {41}},\
  \bibinfo {pages} {5088} (\bibinfo {year} {1990})}\BibitemShut {NoStop}%
\bibitem [{\citenamefont {Landes}\ \emph
  {et~al.}(2021{\natexlab{a}})\citenamefont {Landes}, \citenamefont {Allgaier},
  \citenamefont {Merkouche}, \citenamefont {Smith}, \citenamefont {Marcus},\
  and\ \citenamefont {Raymer}}]{landes2021experimental}%
  \BibitemOpen
  \bibfield  {author} {\bibinfo {author} {\bibfnamefont {T.}~\bibnamefont
  {Landes}}, \bibinfo {author} {\bibfnamefont {M.}~\bibnamefont {Allgaier}},
  \bibinfo {author} {\bibfnamefont {S.}~\bibnamefont {Merkouche}}, \bibinfo
  {author} {\bibfnamefont {B.~J.}\ \bibnamefont {Smith}}, \bibinfo {author}
  {\bibfnamefont {A.~H.}\ \bibnamefont {Marcus}}, \ and\ \bibinfo {author}
  {\bibfnamefont {M.~G.}\ \bibnamefont {Raymer}},\ }\href {\doibase
  10.1103/PhysRevResearch.3.033154} {\bibfield  {journal} {\bibinfo  {journal}
  {Phys. Rev. Research}\ }\textbf {\bibinfo {volume} {3}},\ \bibinfo {pages}
  {033154} (\bibinfo {year} {2021}{\natexlab{a}})}\BibitemShut {NoStop}%
\bibitem [{\citenamefont {Muthukrishnan}\ \emph {et~al.}(2004)\citenamefont
  {Muthukrishnan}, \citenamefont {Agarwal},\ and\ \citenamefont
  {Scully}}]{muthukrishnan2004inducing}%
  \BibitemOpen
  \bibfield  {author} {\bibinfo {author} {\bibfnamefont {A.}~\bibnamefont
  {Muthukrishnan}}, \bibinfo {author} {\bibfnamefont {G.~S.}\ \bibnamefont
  {Agarwal}}, \ and\ \bibinfo {author} {\bibfnamefont {M.~O.}\ \bibnamefont
  {Scully}},\ }\href {\doibase 10.1103/PhysRevLett.93.093002} {\bibfield
  {journal} {\bibinfo  {journal} {Phys. Rev. Lett.}\ }\textbf {\bibinfo
  {volume} {93}},\ \bibinfo {pages} {093002} (\bibinfo {year}
  {2004})}\BibitemShut {NoStop}%
\bibitem [{\citenamefont {Roslyak}\ and\ \citenamefont
  {Mukamel}(2009)}]{roslyak2009multidimensional}%
  \BibitemOpen
  \bibfield  {author} {\bibinfo {author} {\bibfnamefont {O.}~\bibnamefont
  {Roslyak}}\ and\ \bibinfo {author} {\bibfnamefont {S.}~\bibnamefont
  {Mukamel}},\ }\href {\doibase 10.1103/PhysRevA.79.063409} {\bibfield
  {journal} {\bibinfo  {journal} {Phys. Rev. A}\ }\textbf {\bibinfo {volume}
  {79}},\ \bibinfo {pages} {063409} (\bibinfo {year} {2009})}\BibitemShut
  {NoStop}%
\bibitem [{\citenamefont {Schlawin}\ \emph
  {et~al.}(2013{\natexlab{a}})\citenamefont {Schlawin}, \citenamefont
  {Dorfman}, \citenamefont {Fingerhut},\ and\ \citenamefont
  {Mukamel}}]{schlawin2013Suppression}%
  \BibitemOpen
  \bibfield  {author} {\bibinfo {author} {\bibfnamefont {F.}~\bibnamefont
  {Schlawin}}, \bibinfo {author} {\bibfnamefont {K.~E.}\ \bibnamefont
  {Dorfman}}, \bibinfo {author} {\bibfnamefont {B.~P.}\ \bibnamefont
  {Fingerhut}}, \ and\ \bibinfo {author} {\bibfnamefont {S.}~\bibnamefont
  {Mukamel}},\ }\href@noop {} {\bibfield  {journal} {\bibinfo  {journal}
  {Nature Communications}\ }\textbf {\bibinfo {volume} {4}},\ \bibinfo {pages}
  {1782} (\bibinfo {year} {2013}{\natexlab{a}})}\BibitemShut {NoStop}%
\bibitem [{\citenamefont {Shapiro}\ and\ \citenamefont
  {Brumer}(2011)}]{shapiro2011generation}%
  \BibitemOpen
  \bibfield  {author} {\bibinfo {author} {\bibfnamefont {M.}~\bibnamefont
  {Shapiro}}\ and\ \bibinfo {author} {\bibfnamefont {P.}~\bibnamefont
  {Brumer}},\ }\href {\doibase 10.1103/PhysRevLett.106.150501} {\bibfield
  {journal} {\bibinfo  {journal} {Phys. Rev. Lett.}\ }\textbf {\bibinfo
  {volume} {106}},\ \bibinfo {pages} {150501} (\bibinfo {year}
  {2011})}\BibitemShut {NoStop}%
\bibitem [{\citenamefont {Saleh}\ \emph {et~al.}(1998)\citenamefont {Saleh},
  \citenamefont {Jost}, \citenamefont {Fei},\ and\ \citenamefont
  {Teich}}]{saleh}%
  \BibitemOpen
  \bibfield  {author} {\bibinfo {author} {\bibfnamefont {B.~E.~A.}\
  \bibnamefont {Saleh}}, \bibinfo {author} {\bibfnamefont {B.~M.}\ \bibnamefont
  {Jost}}, \bibinfo {author} {\bibfnamefont {H.-B.}\ \bibnamefont {Fei}}, \
  and\ \bibinfo {author} {\bibfnamefont {M.~C.}\ \bibnamefont {Teich}},\ }\href
  {\doibase 10.1103/PhysRevLett.80.3483} {\bibfield  {journal} {\bibinfo
  {journal} {Phys. Rev. Lett.}\ }\textbf {\bibinfo {volume} {80}},\ \bibinfo
  {pages} {3483} (\bibinfo {year} {1998})}\BibitemShut {NoStop}%
\bibitem [{\citenamefont {Kojima}\ and\ \citenamefont {Nguyen}(2004)}]{KOJIMA}%
  \BibitemOpen
  \bibfield  {author} {\bibinfo {author} {\bibfnamefont {J.}~\bibnamefont
  {Kojima}}\ and\ \bibinfo {author} {\bibfnamefont {Q.-V.}\ \bibnamefont
  {Nguyen}},\ }\href {\doibase https://doi.org/10.1016/j.cplett.2004.08.051}
  {\bibfield  {journal} {\bibinfo  {journal} {Chemical Physics Letters}\
  }\textbf {\bibinfo {volume} {396}},\ \bibinfo {pages} {323 } (\bibinfo {year}
  {2004})}\BibitemShut {NoStop}%
\bibitem [{\citenamefont {Pe\ifmmode~\check{r}\else \v{r}\fi{}ina}\ \emph
  {et~al.}(1998)\citenamefont {Pe\ifmmode~\check{r}\else \v{r}\fi{}ina},
  \citenamefont {Saleh},\ and\ \citenamefont {Teich}}]{nphoton}%
  \BibitemOpen
  \bibfield  {author} {\bibinfo {author} {\bibfnamefont {J.}~\bibnamefont
  {Pe\ifmmode~\check{r}\else \v{r}\fi{}ina}}, \bibinfo {author} {\bibfnamefont
  {B.~E.~A.}\ \bibnamefont {Saleh}}, \ and\ \bibinfo {author} {\bibfnamefont
  {M.~C.}\ \bibnamefont {Teich}},\ }\href {\doibase 10.1103/PhysRevA.57.3972}
  {\bibfield  {journal} {\bibinfo  {journal} {Phys. Rev. A}\ }\textbf {\bibinfo
  {volume} {57}},\ \bibinfo {pages} {3972} (\bibinfo {year}
  {1998})}\BibitemShut {NoStop}%
\bibitem [{\citenamefont {de~J.~Le{\'{o}}n-Montiel}\ \emph
  {et~al.}(2013)\citenamefont {de~J.~Le{\'{o}}n-Montiel}, \citenamefont
  {Svozil{\'{\i}}k}, \citenamefont {Salazar-Serrano},\ and\ \citenamefont
  {Torres}}]{roberto_spectral_shape}%
  \BibitemOpen
  \bibfield  {author} {\bibinfo {author} {\bibfnamefont {R.}~\bibnamefont
  {de~J.~Le{\'{o}}n-Montiel}}, \bibinfo {author} {\bibfnamefont
  {J.}~\bibnamefont {Svozil{\'{\i}}k}}, \bibinfo {author} {\bibfnamefont
  {L.~J.}\ \bibnamefont {Salazar-Serrano}}, \ and\ \bibinfo {author}
  {\bibfnamefont {J.~P.}\ \bibnamefont {Torres}},\ }\href {\doibase
  10.1088/1367-2630/15/5/053023} {\bibfield  {journal} {\bibinfo  {journal}
  {New Journal of Physics}\ }\textbf {\bibinfo {volume} {15}},\ \bibinfo
  {pages} {053023} (\bibinfo {year} {2013})}\BibitemShut {NoStop}%
\bibitem [{\citenamefont {Dorfman}\ \emph {et~al.}(2016)\citenamefont
  {Dorfman}, \citenamefont {Schlawin},\ and\ \citenamefont
  {Mukamel}}]{dorfman2016}%
  \BibitemOpen
  \bibfield  {author} {\bibinfo {author} {\bibfnamefont {K.~E.}\ \bibnamefont
  {Dorfman}}, \bibinfo {author} {\bibfnamefont {F.}~\bibnamefont {Schlawin}}, \
  and\ \bibinfo {author} {\bibfnamefont {S.}~\bibnamefont {Mukamel}},\ }\href
  {\doibase 10.1103/RevModPhys.88.045008} {\bibfield  {journal} {\bibinfo
  {journal} {Rev. Mod. Phys.}\ }\textbf {\bibinfo {volume} {88}},\ \bibinfo
  {pages} {045008} (\bibinfo {year} {2016})}\BibitemShut {NoStop}%
\bibitem [{\citenamefont {Schlawin}(2017)}]{schlawin2017}%
  \BibitemOpen
  \bibfield  {author} {\bibinfo {author} {\bibfnamefont {F.}~\bibnamefont
  {Schlawin}},\ }\href@noop {} {\bibfield  {journal} {\bibinfo  {journal} {J.
  Phys. B}\ }\textbf {\bibinfo {volume} {50}},\ \bibinfo {pages} {203001}
  (\bibinfo {year} {2017})}\BibitemShut {NoStop}%
\bibitem [{\citenamefont {Oka}(2010)}]{oka2010}%
  \BibitemOpen
  \bibfield  {author} {\bibinfo {author} {\bibfnamefont {H.}~\bibnamefont
  {Oka}},\ }\href {\doibase 10.1103/PhysRevA.81.063819} {\bibfield  {journal}
  {\bibinfo  {journal} {Phys. Rev. A}\ }\textbf {\bibinfo {volume} {81}},\
  \bibinfo {pages} {063819} (\bibinfo {year} {2010})}\BibitemShut {NoStop}%
\bibitem [{\citenamefont {Schlawin}\ \emph {et~al.}(2018)\citenamefont
  {Schlawin}, \citenamefont {Dorfman},\ and\ \citenamefont
  {Mukamel}}]{Schlawin-2018}%
  \BibitemOpen
  \bibfield  {author} {\bibinfo {author} {\bibfnamefont {F.}~\bibnamefont
  {Schlawin}}, \bibinfo {author} {\bibfnamefont {K.~E.}\ \bibnamefont
  {Dorfman}}, \ and\ \bibinfo {author} {\bibfnamefont {S.}~\bibnamefont
  {Mukamel}},\ }\href {\doibase 10.1021/acs.accounts.8b00173} {\bibfield
  {journal} {\bibinfo  {journal} {Accounts of Chemical Research}\ }\textbf
  {\bibinfo {volume} {51}},\ \bibinfo {pages} {2207} (\bibinfo {year}
  {2018})}\BibitemShut {NoStop}%
\bibitem [{\citenamefont {Villabona-Monsalve}\ \emph
  {et~al.}(2017)\citenamefont {Villabona-Monsalve}, \citenamefont
  {Calderón-Losada}, \citenamefont {Nuñez~Portela},\ and\ \citenamefont
  {Valencia}}]{villabona_calderon_2017}%
  \BibitemOpen
  \bibfield  {author} {\bibinfo {author} {\bibfnamefont {J.~P.}\ \bibnamefont
  {Villabona-Monsalve}}, \bibinfo {author} {\bibfnamefont {O.}~\bibnamefont
  {Calderón-Losada}}, \bibinfo {author} {\bibfnamefont {M.}~\bibnamefont
  {Nuñez~Portela}}, \ and\ \bibinfo {author} {\bibfnamefont {A.}~\bibnamefont
  {Valencia}},\ }\href {\doibase 10.1021/acs.jpca.7b06450} {\bibfield
  {journal} {\bibinfo  {journal} {The Journal of Physical Chemistry A}\
  }\textbf {\bibinfo {volume} {121}},\ \bibinfo {pages} {7869} (\bibinfo {year}
  {2017})}\BibitemShut {NoStop}%
\bibitem [{\citenamefont {Varnavski}\ \emph {et~al.}(2017)\citenamefont
  {Varnavski}, \citenamefont {Pinsky},\ and\ \citenamefont
  {Goodson}}]{Varnavski2017}%
  \BibitemOpen
  \bibfield  {author} {\bibinfo {author} {\bibfnamefont {O.}~\bibnamefont
  {Varnavski}}, \bibinfo {author} {\bibfnamefont {B.}~\bibnamefont {Pinsky}}, \
  and\ \bibinfo {author} {\bibfnamefont {T.}~\bibnamefont {Goodson}},\ }\href
  {\doibase 10.1021/acs.jpclett.6b02378} {\bibfield  {journal} {\bibinfo
  {journal} {The Journal of Physical Chemistry Letters}\ }\textbf {\bibinfo
  {volume} {8}},\ \bibinfo {pages} {388} (\bibinfo {year} {2017})}\BibitemShut
  {NoStop}%
\bibitem [{\citenamefont {Oka}(2018{\natexlab{a}})}]{oka2018-1}%
  \BibitemOpen
  \bibfield  {author} {\bibinfo {author} {\bibfnamefont {H.}~\bibnamefont
  {Oka}},\ }\href {\doibase 10.1103/PhysRevA.97.063859} {\bibfield  {journal}
  {\bibinfo  {journal} {Phys. Rev. A}\ }\textbf {\bibinfo {volume} {97}},\
  \bibinfo {pages} {063859} (\bibinfo {year} {2018}{\natexlab{a}})}\BibitemShut
  {NoStop}%
\bibitem [{\citenamefont {Oka}(2018{\natexlab{b}})}]{oka2018-2}%
  \BibitemOpen
  \bibfield  {author} {\bibinfo {author} {\bibfnamefont {H.}~\bibnamefont
  {Oka}},\ }\href {\doibase 10.1103/PhysRevA.97.033814} {\bibfield  {journal}
  {\bibinfo  {journal} {Phys. Rev. A}\ }\textbf {\bibinfo {volume} {97}},\
  \bibinfo {pages} {033814} (\bibinfo {year} {2018}{\natexlab{b}})}\BibitemShut
  {NoStop}%
\bibitem [{\citenamefont {Svozil\'{i}k}\ \emph {et~al.}(2018)\citenamefont
  {Svozil\'{i}k}, \citenamefont {Pe\v{r}ina},\ and\ \citenamefont
  {de~J.~Le\'{o}n-Montiel}}]{svozilik2018-1}%
  \BibitemOpen
  \bibfield  {author} {\bibinfo {author} {\bibfnamefont {J.}~\bibnamefont
  {Svozil\'{i}k}}, \bibinfo {author} {\bibfnamefont {J.}~\bibnamefont
  {Pe\v{r}ina}}, \ and\ \bibinfo {author} {\bibfnamefont {R.}~\bibnamefont
  {de~J.~Le\'{o}n-Montiel}},\ }\href {\doibase 10.1364/JOSAB.35.000460}
  {\bibfield  {journal} {\bibinfo  {journal} {J. Opt. Soc. Am. B}\ }\textbf
  {\bibinfo {volume} {35}},\ \bibinfo {pages} {460} (\bibinfo {year}
  {2018})}\BibitemShut {NoStop}%
\bibitem [{\citenamefont {Svozilík}\ \emph {et~al.}(2018)\citenamefont
  {Svozilík}, \citenamefont {Peřina},\ and\ \citenamefont
  {de~J.~Le{\'{o}}n-Montiel}}]{svozilik2018-2}%
  \BibitemOpen
  \bibfield  {author} {\bibinfo {author} {\bibfnamefont {J.}~\bibnamefont
  {Svozilík}}, \bibinfo {author} {\bibfnamefont {J.}~\bibnamefont {Peřina}},
  \ and\ \bibinfo {author} {\bibfnamefont {R.}~\bibnamefont
  {de~J.~Le{\'{o}}n-Montiel}},\ }\href {\doibase
  https://doi.org/10.1016/j.chemphys.2018.05.013} {\bibfield  {journal}
  {\bibinfo  {journal} {Chemical Physics}\ }\textbf {\bibinfo {volume} {510}},\
  \bibinfo {pages} {54} (\bibinfo {year} {2018})}\BibitemShut {NoStop}%
\bibitem [{\citenamefont {Burdick}\ \emph {et~al.}(2018)\citenamefont
  {Burdick}, \citenamefont {Varnavski}, \citenamefont {Molina}, \citenamefont
  {Upton}, \citenamefont {Zimmerman},\ and\ \citenamefont
  {Goodson}}]{burdick2018}%
  \BibitemOpen
  \bibfield  {author} {\bibinfo {author} {\bibfnamefont {R.~K.}\ \bibnamefont
  {Burdick}}, \bibinfo {author} {\bibfnamefont {O.}~\bibnamefont {Varnavski}},
  \bibinfo {author} {\bibfnamefont {A.}~\bibnamefont {Molina}}, \bibinfo
  {author} {\bibfnamefont {L.}~\bibnamefont {Upton}}, \bibinfo {author}
  {\bibfnamefont {P.}~\bibnamefont {Zimmerman}}, \ and\ \bibinfo {author}
  {\bibfnamefont {T.}~\bibnamefont {Goodson}},\ }\href {\doibase
  10.1021/acs.jpca.8b07466} {\bibfield  {journal} {\bibinfo  {journal} {The
  Journal of Physical Chemistry A}\ }\textbf {\bibinfo {volume} {122}},\
  \bibinfo {pages} {8198} (\bibinfo {year} {2018})}\BibitemShut {NoStop}%
\bibitem [{\citenamefont {de~J.~Le{\'{o}}n-Montiel}\ \emph
  {et~al.}(2019)\citenamefont {de~J.~Le{\'{o}}n-Montiel}, \citenamefont
  {Svozil\'{\i}k}, \citenamefont {Torres},\ and\ \citenamefont
  {U'Ren}}]{RobertoTemperatureControlled}%
  \BibitemOpen
  \bibfield  {author} {\bibinfo {author} {\bibfnamefont {R.}~\bibnamefont
  {de~J.~Le{\'{o}}n-Montiel}}, \bibinfo {author} {\bibfnamefont
  {J.}~\bibnamefont {Svozil\'{\i}k}}, \bibinfo {author} {\bibfnamefont {J.~P.}\
  \bibnamefont {Torres}}, \ and\ \bibinfo {author} {\bibfnamefont {A.~B.}\
  \bibnamefont {U'Ren}},\ }\href {\doibase 10.1103/PhysRevLett.123.023601}
  {\bibfield  {journal} {\bibinfo  {journal} {Phys. Rev. Lett.}\ }\textbf
  {\bibinfo {volume} {123}},\ \bibinfo {pages} {023601} (\bibinfo {year}
  {2019})}\BibitemShut {NoStop}%
\bibitem [{\citenamefont {Mukamel}\ \emph {et~al.}(2020)\citenamefont
  {Mukamel}, \citenamefont {Freyberger}, \citenamefont {Schleich},
  \citenamefont {Bellini}, \citenamefont {Zavatta}, \citenamefont {Leuchs},
  \citenamefont {Silberhorn}, \citenamefont {Boyd}, \citenamefont
  {S{\'{a}}nchez-Soto}, \citenamefont {Stefanov}, \citenamefont {Barbieri},
  \citenamefont {Paterova}, \citenamefont {Krivitsky}, \citenamefont {Shwartz},
  \citenamefont {Tamasaku}, \citenamefont {Dorfman}, \citenamefont {Schlawin},
  \citenamefont {Sandoghdar}, \citenamefont {Raymer}, \citenamefont {Marcus},
  \citenamefont {Varnavski}, \citenamefont {Goodson}, \citenamefont {Zhou},
  \citenamefont {Shi}, \citenamefont {Asban}, \citenamefont {Scully},
  \citenamefont {Agarwal}, \citenamefont {Peng}, \citenamefont {Sokolov},
  \citenamefont {Zhang}, \citenamefont {Zubairy}, \citenamefont {Vartanyants},
  \citenamefont {del Valle},\ and\ \citenamefont
  {Laussy}}]{Mukamel2020roadmap}%
  \BibitemOpen
  \bibfield  {author} {\bibinfo {author} {\bibfnamefont {S.}~\bibnamefont
  {Mukamel}}, \bibinfo {author} {\bibfnamefont {M.}~\bibnamefont {Freyberger}},
  \bibinfo {author} {\bibfnamefont {W.}~\bibnamefont {Schleich}}, \bibinfo
  {author} {\bibfnamefont {M.}~\bibnamefont {Bellini}}, \bibinfo {author}
  {\bibfnamefont {A.}~\bibnamefont {Zavatta}}, \bibinfo {author} {\bibfnamefont
  {G.}~\bibnamefont {Leuchs}}, \bibinfo {author} {\bibfnamefont
  {C.}~\bibnamefont {Silberhorn}}, \bibinfo {author} {\bibfnamefont {R.~W.}\
  \bibnamefont {Boyd}}, \bibinfo {author} {\bibfnamefont {L.~L.}\ \bibnamefont
  {S{\'{a}}nchez-Soto}}, \bibinfo {author} {\bibfnamefont {A.}~\bibnamefont
  {Stefanov}}, \bibinfo {author} {\bibfnamefont {M.}~\bibnamefont {Barbieri}},
  \bibinfo {author} {\bibfnamefont {A.}~\bibnamefont {Paterova}}, \bibinfo
  {author} {\bibfnamefont {L.}~\bibnamefont {Krivitsky}}, \bibinfo {author}
  {\bibfnamefont {S.}~\bibnamefont {Shwartz}}, \bibinfo {author} {\bibfnamefont
  {K.}~\bibnamefont {Tamasaku}}, \bibinfo {author} {\bibfnamefont
  {K.}~\bibnamefont {Dorfman}}, \bibinfo {author} {\bibfnamefont
  {F.}~\bibnamefont {Schlawin}}, \bibinfo {author} {\bibfnamefont
  {V.}~\bibnamefont {Sandoghdar}}, \bibinfo {author} {\bibfnamefont
  {M.}~\bibnamefont {Raymer}}, \bibinfo {author} {\bibfnamefont
  {A.}~\bibnamefont {Marcus}}, \bibinfo {author} {\bibfnamefont
  {O.}~\bibnamefont {Varnavski}}, \bibinfo {author} {\bibfnamefont
  {T.}~\bibnamefont {Goodson}}, \bibinfo {author} {\bibfnamefont {Z.-Y.}\
  \bibnamefont {Zhou}}, \bibinfo {author} {\bibfnamefont {B.-S.}\ \bibnamefont
  {Shi}}, \bibinfo {author} {\bibfnamefont {S.}~\bibnamefont {Asban}}, \bibinfo
  {author} {\bibfnamefont {M.}~\bibnamefont {Scully}}, \bibinfo {author}
  {\bibfnamefont {G.}~\bibnamefont {Agarwal}}, \bibinfo {author} {\bibfnamefont
  {T.}~\bibnamefont {Peng}}, \bibinfo {author} {\bibfnamefont {A.~V.}\
  \bibnamefont {Sokolov}}, \bibinfo {author} {\bibfnamefont {Z.-D.}\
  \bibnamefont {Zhang}}, \bibinfo {author} {\bibfnamefont {M.~S.}\ \bibnamefont
  {Zubairy}}, \bibinfo {author} {\bibfnamefont {I.~A.}\ \bibnamefont
  {Vartanyants}}, \bibinfo {author} {\bibfnamefont {E.}~\bibnamefont {del
  Valle}}, \ and\ \bibinfo {author} {\bibfnamefont {F.}~\bibnamefont
  {Laussy}},\ }\href {\doibase 10.1088/1361-6455/ab69a8} {\bibfield  {journal}
  {\bibinfo  {journal} {Journal of Physics B: Atomic, Molecular and Optical
  Physics}\ }\textbf {\bibinfo {volume} {53}},\ \bibinfo {pages} {072002}
  (\bibinfo {year} {2020})}\BibitemShut {NoStop}%
\bibitem [{\citenamefont {Raymer}\ \emph
  {et~al.}(2021{\natexlab{a}})\citenamefont {Raymer}, \citenamefont {Landes},\
  and\ \citenamefont {Marcus}}]{raymer2021entangled}%
  \BibitemOpen
  \bibfield  {author} {\bibinfo {author} {\bibfnamefont {M.~G.}\ \bibnamefont
  {Raymer}}, \bibinfo {author} {\bibfnamefont {T.}~\bibnamefont {Landes}}, \
  and\ \bibinfo {author} {\bibfnamefont {A.~H.}\ \bibnamefont {Marcus}},\
  }\href {\doibase 10.1063/5.0049338} {\bibfield  {journal} {\bibinfo
  {journal} {The Journal of Chemical Physics}\ }\textbf {\bibinfo {volume}
  {155}},\ \bibinfo {pages} {081501} (\bibinfo {year}
  {2021}{\natexlab{a}})}\BibitemShut {NoStop}%
\bibitem [{\citenamefont {Landes}\ \emph
  {et~al.}(2021{\natexlab{b}})\citenamefont {Landes}, \citenamefont {Raymer},
  \citenamefont {Allgaier}, \citenamefont {Merkouche}, \citenamefont {Smith},\
  and\ \citenamefont {Marcus}}]{landes2021quantifying}%
  \BibitemOpen
  \bibfield  {author} {\bibinfo {author} {\bibfnamefont {T.}~\bibnamefont
  {Landes}}, \bibinfo {author} {\bibfnamefont {M.~G.}\ \bibnamefont {Raymer}},
  \bibinfo {author} {\bibfnamefont {M.}~\bibnamefont {Allgaier}}, \bibinfo
  {author} {\bibfnamefont {S.}~\bibnamefont {Merkouche}}, \bibinfo {author}
  {\bibfnamefont {B.~J.}\ \bibnamefont {Smith}}, \ and\ \bibinfo {author}
  {\bibfnamefont {A.~H.}\ \bibnamefont {Marcus}},\ }\href {\doibase
  10.1364/OE.422544} {\bibfield  {journal} {\bibinfo  {journal} {Opt. Express}\
  }\textbf {\bibinfo {volume} {29}},\ \bibinfo {pages} {20022} (\bibinfo {year}
  {2021}{\natexlab{b}})}\BibitemShut {NoStop}%
\bibitem [{\citenamefont {Raymer}\ \emph
  {et~al.}(2021{\natexlab{b}})\citenamefont {Raymer}, \citenamefont {Landes},
  \citenamefont {Allgaier}, \citenamefont {Merkouche}, \citenamefont {Smith},\
  and\ \citenamefont {Marcus}}]{raymer2021}%
  \BibitemOpen
  \bibfield  {author} {\bibinfo {author} {\bibfnamefont {M.~G.}\ \bibnamefont
  {Raymer}}, \bibinfo {author} {\bibfnamefont {T.}~\bibnamefont {Landes}},
  \bibinfo {author} {\bibfnamefont {M.}~\bibnamefont {Allgaier}}, \bibinfo
  {author} {\bibfnamefont {S.}~\bibnamefont {Merkouche}}, \bibinfo {author}
  {\bibfnamefont {B.~J.}\ \bibnamefont {Smith}}, \ and\ \bibinfo {author}
  {\bibfnamefont {A.~H.}\ \bibnamefont {Marcus}},\ }\href {\doibase
  10.1364/OPTICA.426674} {\bibfield  {journal} {\bibinfo  {journal} {Optica}\
  }\textbf {\bibinfo {volume} {8}},\ \bibinfo {pages} {757} (\bibinfo {year}
  {2021}{\natexlab{b}})}\BibitemShut {NoStop}%
\bibitem [{\citenamefont {Villabona-Monsalve}\ \emph
  {et~al.}(2020)\citenamefont {Villabona-Monsalve}, \citenamefont {Burdick},\
  and\ \citenamefont {Goodson}}]{villabona2020}%
  \BibitemOpen
  \bibfield  {author} {\bibinfo {author} {\bibfnamefont {J.~P.}\ \bibnamefont
  {Villabona-Monsalve}}, \bibinfo {author} {\bibfnamefont {R.~K.}\ \bibnamefont
  {Burdick}}, \ and\ \bibinfo {author} {\bibfnamefont {T.}~\bibnamefont
  {Goodson}},\ }\href {\doibase 10.1021/acs.jpcc.0c08678} {\bibfield  {journal}
  {\bibinfo  {journal} {The Journal of Physical Chemistry C}\ }\textbf
  {\bibinfo {volume} {124}},\ \bibinfo {pages} {24526} (\bibinfo {year}
  {2020})}\BibitemShut {NoStop}%
\bibitem [{\citenamefont {Parzuchowski}\ \emph {et~al.}(2021)\citenamefont
  {Parzuchowski}, \citenamefont {Mikhaylov}, \citenamefont {Mazurek},
  \citenamefont {Wilson}, \citenamefont {Lum}, \citenamefont {Gerrits},
  \citenamefont {Camp}, \citenamefont {Stevens},\ and\ \citenamefont
  {Jimenez}}]{parzuchowski2021}%
  \BibitemOpen
  \bibfield  {author} {\bibinfo {author} {\bibfnamefont {K.~M.}\ \bibnamefont
  {Parzuchowski}}, \bibinfo {author} {\bibfnamefont {A.}~\bibnamefont
  {Mikhaylov}}, \bibinfo {author} {\bibfnamefont {M.~D.}\ \bibnamefont
  {Mazurek}}, \bibinfo {author} {\bibfnamefont {R.~N.}\ \bibnamefont {Wilson}},
  \bibinfo {author} {\bibfnamefont {D.~J.}\ \bibnamefont {Lum}}, \bibinfo
  {author} {\bibfnamefont {T.}~\bibnamefont {Gerrits}}, \bibinfo {author}
  {\bibfnamefont {C.~H.}\ \bibnamefont {Camp}}, \bibinfo {author}
  {\bibfnamefont {M.~J.}\ \bibnamefont {Stevens}}, \ and\ \bibinfo {author}
  {\bibfnamefont {R.}~\bibnamefont {Jimenez}},\ }\href {\doibase
  10.1103/PhysRevApplied.15.044012} {\bibfield  {journal} {\bibinfo  {journal}
  {Phys. Rev. Applied}\ }\textbf {\bibinfo {volume} {15}},\ \bibinfo {pages}
  {044012} (\bibinfo {year} {2021})}\BibitemShut {NoStop}%
\bibitem [{\citenamefont {Tabakaev}\ \emph {et~al.}(2021)\citenamefont
  {Tabakaev}, \citenamefont {Montagnese}, \citenamefont {Haack}, \citenamefont
  {Bonacina}, \citenamefont {Wolf}, \citenamefont {Zbinden},\ and\
  \citenamefont {Thew}}]{tabakaev2021}%
  \BibitemOpen
  \bibfield  {author} {\bibinfo {author} {\bibfnamefont {D.}~\bibnamefont
  {Tabakaev}}, \bibinfo {author} {\bibfnamefont {M.}~\bibnamefont
  {Montagnese}}, \bibinfo {author} {\bibfnamefont {G.}~\bibnamefont {Haack}},
  \bibinfo {author} {\bibfnamefont {L.}~\bibnamefont {Bonacina}}, \bibinfo
  {author} {\bibfnamefont {J.-P.}\ \bibnamefont {Wolf}}, \bibinfo {author}
  {\bibfnamefont {H.}~\bibnamefont {Zbinden}}, \ and\ \bibinfo {author}
  {\bibfnamefont {R.~T.}\ \bibnamefont {Thew}},\ }\href {\doibase
  10.1103/PhysRevA.103.033701} {\bibfield  {journal} {\bibinfo  {journal}
  {Phys. Rev. A}\ }\textbf {\bibinfo {volume} {103}},\ \bibinfo {pages}
  {033701} (\bibinfo {year} {2021})}\BibitemShut {NoStop}%
\bibitem [{\citenamefont {Corona-Aquino}\ \emph {et~al.}(2021)\citenamefont
  {Corona-Aquino}, \citenamefont {Calder\'{o}n-Losada}, \citenamefont
  {Li-G\'{o}mez}, \citenamefont {Cruz-Ram\'{i}rez}, \citenamefont
  {Alvarez-Venicio}, \citenamefont {Carre\'{o}n-Castro}, \citenamefont
  {de~J.~Le{\'{o}}n-Montiel},\ and\ \citenamefont {U'Ren}}]{samuel2021}%
  \BibitemOpen
  \bibfield  {author} {\bibinfo {author} {\bibfnamefont {S.}~\bibnamefont
  {Corona-Aquino}}, \bibinfo {author} {\bibfnamefont {O.}~\bibnamefont
  {Calder\'{o}n-Losada}}, \bibinfo {author} {\bibfnamefont {M.~Y.}\
  \bibnamefont {Li-G\'{o}mez}}, \bibinfo {author} {\bibfnamefont
  {H.}~\bibnamefont {Cruz-Ram\'{i}rez}}, \bibinfo {author} {\bibfnamefont
  {V.}~\bibnamefont {Alvarez-Venicio}}, \bibinfo {author} {\bibfnamefont
  {M.~P.}\ \bibnamefont {Carre\'{o}n-Castro}}, \bibinfo {author} {\bibfnamefont
  {R.}~\bibnamefont {de~J.~Le{\'{o}}n-Montiel}}, \ and\ \bibinfo {author}
  {\bibfnamefont {A.~B.}\ \bibnamefont {U'Ren}},\ }\href@noop {} {\bibfield
  {journal} {\bibinfo  {journal} {arXiv:2101.10987}\ } (\bibinfo {year}
  {2021})}\BibitemShut {NoStop}%
\bibitem [{\citenamefont {Schlawin}\ \emph
  {et~al.}(2013{\natexlab{b}})\citenamefont {Schlawin}, \citenamefont
  {Dorfman}, \citenamefont {Fingerhut},\ and\ \citenamefont
  {Mukamel}}]{paper2}%
  \BibitemOpen
  \bibfield  {author} {\bibinfo {author} {\bibfnamefont {F.}~\bibnamefont
  {Schlawin}}, \bibinfo {author} {\bibfnamefont {K.}~\bibnamefont {Dorfman}},
  \bibinfo {author} {\bibfnamefont {B.}~\bibnamefont {Fingerhut}}, \ and\
  \bibinfo {author} {\bibfnamefont {S.}~\bibnamefont {Mukamel}},\ }\href
  {\doibase 10.1038/ncomms2802} {\bibfield  {journal} {\bibinfo  {journal}
  {Nature communications}\ }\textbf {\bibinfo {volume} {4}},\ \bibinfo {pages}
  {1782} (\bibinfo {year} {2013}{\natexlab{b}})}\BibitemShut {NoStop}%
\bibitem [{\citenamefont {Hipke}\ \emph {et~al.}(2014)\citenamefont {Hipke},
  \citenamefont {Meek}, \citenamefont {Ideguchi}, \citenamefont {H\"ansch},\
  and\ \citenamefont {Picqu\'e}}]{hipke2014broadband}%
  \BibitemOpen
  \bibfield  {author} {\bibinfo {author} {\bibfnamefont {A.}~\bibnamefont
  {Hipke}}, \bibinfo {author} {\bibfnamefont {S.~A.}\ \bibnamefont {Meek}},
  \bibinfo {author} {\bibfnamefont {T.}~\bibnamefont {Ideguchi}}, \bibinfo
  {author} {\bibfnamefont {T.~W.}\ \bibnamefont {H\"ansch}}, \ and\ \bibinfo
  {author} {\bibfnamefont {N.}~\bibnamefont {Picqu\'e}},\ }\href {\doibase
  10.1103/PhysRevA.90.011805} {\bibfield  {journal} {\bibinfo  {journal} {Phys.
  Rev. A}\ }\textbf {\bibinfo {volume} {90}},\ \bibinfo {pages} {011805}
  (\bibinfo {year} {2014})}\BibitemShut {NoStop}%
\bibitem [{\citenamefont {Erling}\ \emph {et~al.}(2019)\citenamefont {Erling},
  \citenamefont {Thyrhaug}, \citenamefont {Stefan}, \citenamefont {Krause},
  \citenamefont {Antonio}, \citenamefont {Perri}, \citenamefont {Giulio},
  \citenamefont {Cerullo}, \citenamefont {Dario},\ and\ \citenamefont
  {Polli}}]{Erling2019Single}%
  \BibitemOpen
  \bibfield  {author} {\bibinfo {author} {\bibnamefont {Erling}}, \bibinfo
  {author} {\bibnamefont {Thyrhaug}}, \bibinfo {author} {\bibnamefont
  {Stefan}}, \bibinfo {author} {\bibnamefont {Krause}}, \bibinfo {author}
  {\bibnamefont {Antonio}}, \bibinfo {author} {\bibnamefont {Perri}}, \bibinfo
  {author} {\bibnamefont {Giulio}}, \bibinfo {author} {\bibnamefont {Cerullo}},
  \bibinfo {author} {\bibnamefont {Dario}}, \ and\ \bibinfo {author}
  {\bibnamefont {Polli}},\ }\href@noop {} {\bibfield  {journal} {\bibinfo
  {journal} {Proceedings of the National Academy of Sciences of the United
  States of America}\ }\textbf {\bibinfo {volume} {116}} (\bibinfo {year}
  {2019})}\BibitemShut {NoStop}%
\bibitem [{\citenamefont {Piatkowski}\ \emph {et~al.}(2016)\citenamefont
  {Piatkowski}, \citenamefont {Gellings},\ and\ \citenamefont
  {Hulst}}]{piatkowski2016Broadband}%
  \BibitemOpen
  \bibfield  {author} {\bibinfo {author} {\bibfnamefont {L.}~\bibnamefont
  {Piatkowski}}, \bibinfo {author} {\bibfnamefont {E.}~\bibnamefont
  {Gellings}}, \ and\ \bibinfo {author} {\bibfnamefont {N.~V.}\ \bibnamefont
  {Hulst}},\ }\href@noop {} {\bibfield  {journal} {\bibinfo  {journal} {Nature
  Communications}\ }\textbf {\bibinfo {volume} {7}},\ \bibinfo {pages} {10411}
  (\bibinfo {year} {2016})}\BibitemShut {NoStop}%
\bibitem [{\citenamefont {Hong}\ \emph {et~al.}(1987)\citenamefont {Hong},
  \citenamefont {Ou},\ and\ \citenamefont {Mandel}}]{hong1987measurement}%
  \BibitemOpen
  \bibfield  {author} {\bibinfo {author} {\bibfnamefont {C.~K.}\ \bibnamefont
  {Hong}}, \bibinfo {author} {\bibfnamefont {Z.~Y.}\ \bibnamefont {Ou}}, \ and\
  \bibinfo {author} {\bibfnamefont {L.}~\bibnamefont {Mandel}},\ }\href
  {\doibase 10.1103/PhysRevLett.59.2044} {\bibfield  {journal} {\bibinfo
  {journal} {Phys. Rev. Lett.}\ }\textbf {\bibinfo {volume} {59}},\ \bibinfo
  {pages} {2044} (\bibinfo {year} {1987})}\BibitemShut {NoStop}%
\bibitem [{\citenamefont {Chou}\ \emph {et~al.}(2020)\citenamefont {Chou},
  \citenamefont {Collopy}, \citenamefont {Kurz}, \citenamefont {Lin},
  \citenamefont {Harding}, \citenamefont {Plessow}, \citenamefont {Fortier},
  \citenamefont {Diddams}, \citenamefont {Leibfried},\ and\ \citenamefont
  {Leibrandt}}]{chou2020Frequency}%
  \BibitemOpen
  \bibfield  {author} {\bibinfo {author} {\bibfnamefont {C.~W.}\ \bibnamefont
  {Chou}}, \bibinfo {author} {\bibfnamefont {A.~L.}\ \bibnamefont {Collopy}},
  \bibinfo {author} {\bibfnamefont {C.}~\bibnamefont {Kurz}}, \bibinfo {author}
  {\bibfnamefont {Y.}~\bibnamefont {Lin}}, \bibinfo {author} {\bibfnamefont
  {M.~E.}\ \bibnamefont {Harding}}, \bibinfo {author} {\bibfnamefont {P.~N.}\
  \bibnamefont {Plessow}}, \bibinfo {author} {\bibfnamefont {T.}~\bibnamefont
  {Fortier}}, \bibinfo {author} {\bibfnamefont {S.}~\bibnamefont {Diddams}},
  \bibinfo {author} {\bibfnamefont {D.}~\bibnamefont {Leibfried}}, \ and\
  \bibinfo {author} {\bibfnamefont {D.~R.}\ \bibnamefont {Leibrandt}},\ }\href
  {\doibase 10.1126/science.aba3628} {\bibfield  {journal} {\bibinfo  {journal}
  {Science}\ }\textbf {\bibinfo {volume} {367}} (\bibinfo {year} {2020}),\
  10.1126/science.aba3628}\BibitemShut {NoStop}%
\bibitem [{\citenamefont {You}\ \emph {et~al.}(2021)\citenamefont {You},
  \citenamefont {Hong}, \citenamefont {Bierhorst}, \citenamefont {Lita},
  \citenamefont {Glancy}, \citenamefont {Kolthammer}, \citenamefont {Knill},
  \citenamefont {Nam}, \citenamefont {Mirin}, \citenamefont {Magaña-Loaiza},\
  and\ \citenamefont {Gerrits}}]{chenglong2021}%
  \BibitemOpen
  \bibfield  {author} {\bibinfo {author} {\bibfnamefont {C.}~\bibnamefont
  {You}}, \bibinfo {author} {\bibfnamefont {M.}~\bibnamefont {Hong}}, \bibinfo
  {author} {\bibfnamefont {P.}~\bibnamefont {Bierhorst}}, \bibinfo {author}
  {\bibfnamefont {A.~E.}\ \bibnamefont {Lita}}, \bibinfo {author}
  {\bibfnamefont {S.}~\bibnamefont {Glancy}}, \bibinfo {author} {\bibfnamefont
  {S.}~\bibnamefont {Kolthammer}}, \bibinfo {author} {\bibfnamefont
  {E.}~\bibnamefont {Knill}}, \bibinfo {author} {\bibfnamefont {S.~W.}\
  \bibnamefont {Nam}}, \bibinfo {author} {\bibfnamefont {R.~P.}\ \bibnamefont
  {Mirin}}, \bibinfo {author} {\bibfnamefont {O.~S.}\ \bibnamefont
  {Magaña-Loaiza}}, \ and\ \bibinfo {author} {\bibfnamefont {T.}~\bibnamefont
  {Gerrits}},\ }\href {\doibase 10.1063/5.0063294} {\bibfield  {journal}
  {\bibinfo  {journal} {Applied Physics Reviews}\ }\textbf {\bibinfo {volume}
  {8}},\ \bibinfo {pages} {041406} (\bibinfo {year} {2021})}\BibitemShut
  {NoStop}%
\bibitem [{\citenamefont {Hong}\ \emph {et~al.}(2021)\citenamefont {Hong},
  \citenamefont {ur~Rehman}, \citenamefont {Kim}, \citenamefont {Cho},
  \citenamefont {Lee}, \citenamefont {Jung}, \citenamefont {Moon},
  \citenamefont {Han},\ and\ \citenamefont {Lim}}]{hong2021quantum}%
  \BibitemOpen
  \bibfield  {author} {\bibinfo {author} {\bibfnamefont {S.}~\bibnamefont
  {Hong}}, \bibinfo {author} {\bibfnamefont {J.}~\bibnamefont {ur~Rehman}},
  \bibinfo {author} {\bibfnamefont {Y.-S.}\ \bibnamefont {Kim}}, \bibinfo
  {author} {\bibfnamefont {Y.-W.}\ \bibnamefont {Cho}}, \bibinfo {author}
  {\bibfnamefont {S.-W.}\ \bibnamefont {Lee}}, \bibinfo {author} {\bibfnamefont
  {H.}~\bibnamefont {Jung}}, \bibinfo {author} {\bibfnamefont {S.}~\bibnamefont
  {Moon}}, \bibinfo {author} {\bibfnamefont {S.-W.}\ \bibnamefont {Han}}, \
  and\ \bibinfo {author} {\bibfnamefont {H.-T.}\ \bibnamefont {Lim}},\ }\href
  {\doibase 10.1038/s41467-021-25451-4} {\bibfield  {journal} {\bibinfo
  {journal} {Nature Communications}\ }\textbf {\bibinfo {volume} {12}}
  (\bibinfo {year} {2021}),\ 10.1038/s41467-021-25451-4}\BibitemShut {NoStop}%
\bibitem [{\citenamefont {Lyons}\ \emph {et~al.}(2018)\citenamefont {Lyons},
  \citenamefont {Knee}, \citenamefont {Bolduc}, \citenamefont {Roger},
  \citenamefont {Leach}, \citenamefont {Gauger},\ and\ \citenamefont
  {Faccio}}]{lyons2018Attosecond}%
  \BibitemOpen
  \bibfield  {author} {\bibinfo {author} {\bibfnamefont {A.}~\bibnamefont
  {Lyons}}, \bibinfo {author} {\bibfnamefont {G.~C.}\ \bibnamefont {Knee}},
  \bibinfo {author} {\bibfnamefont {E.}~\bibnamefont {Bolduc}}, \bibinfo
  {author} {\bibfnamefont {T.}~\bibnamefont {Roger}}, \bibinfo {author}
  {\bibfnamefont {J.}~\bibnamefont {Leach}}, \bibinfo {author} {\bibfnamefont
  {E.~M.}\ \bibnamefont {Gauger}}, \ and\ \bibinfo {author} {\bibfnamefont
  {D.}~\bibnamefont {Faccio}},\ }\href@noop {} {\bibfield  {journal} {\bibinfo
  {journal} {Science Advances}\ }\textbf {\bibinfo {volume} {4}},\ \bibinfo
  {pages} {eaap9416} (\bibinfo {year} {2018})}\BibitemShut {NoStop}%
\bibitem [{\citenamefont {Chen}\ \emph {et~al.}(2019)\citenamefont {Chen},
  \citenamefont {Fink}, \citenamefont {Steinlechner}, \citenamefont {Torres},\
  and\ \citenamefont {Ursin}}]{chen2019Hong}%
  \BibitemOpen
  \bibfield  {author} {\bibinfo {author} {\bibfnamefont {Y.}~\bibnamefont
  {Chen}}, \bibinfo {author} {\bibfnamefont {M.}~\bibnamefont {Fink}}, \bibinfo
  {author} {\bibfnamefont {F.}~\bibnamefont {Steinlechner}}, \bibinfo {author}
  {\bibfnamefont {J.~P.}\ \bibnamefont {Torres}}, \ and\ \bibinfo {author}
  {\bibfnamefont {R.}~\bibnamefont {Ursin}},\ }\href@noop {} {\bibfield
  {journal} {\bibinfo  {journal} {npj Quantum Information}\ }\textbf {\bibinfo
  {volume} {5}},\  (\bibinfo {year} {2019})}\BibitemShut {NoStop}%
\bibitem [{\citenamefont {Hiekkamki}\ \emph {et~al.}(2021)\citenamefont
  {Hiekkamki}, \citenamefont {Bouchard},\ and\ \citenamefont
  {Fickler}}]{hiekkamki2021Photonic}%
  \BibitemOpen
  \bibfield  {author} {\bibinfo {author} {\bibfnamefont {M.}~\bibnamefont
  {Hiekkamki}}, \bibinfo {author} {\bibfnamefont {F.}~\bibnamefont {Bouchard}},
  \ and\ \bibinfo {author} {\bibfnamefont {R.}~\bibnamefont {Fickler}},\
  }\href@noop {} {\enquote {\bibinfo {title} {Photonic angular super-resolution
  using twisted n00n state},}\ } (\bibinfo {year} {2021}),\ \Eprint
  {http://arxiv.org/abs/2106.09273} {arXiv:2106.09273 [quant-ph]} \BibitemShut
  {NoStop}%
\bibitem [{\citenamefont {Roger}\ \emph {et~al.}(2016)\citenamefont {Roger},
  \citenamefont {Restuccia}, \citenamefont {Lyons}, \citenamefont {Giovannini},
  \citenamefont {Romero}, \citenamefont {Jeffers}, \citenamefont {Padgett},\
  and\ \citenamefont {Faccio}}]{roger2016coherent}%
  \BibitemOpen
  \bibfield  {author} {\bibinfo {author} {\bibfnamefont {T.}~\bibnamefont
  {Roger}}, \bibinfo {author} {\bibfnamefont {S.}~\bibnamefont {Restuccia}},
  \bibinfo {author} {\bibfnamefont {A.}~\bibnamefont {Lyons}}, \bibinfo
  {author} {\bibfnamefont {D.}~\bibnamefont {Giovannini}}, \bibinfo {author}
  {\bibfnamefont {J.}~\bibnamefont {Romero}}, \bibinfo {author} {\bibfnamefont
  {J.}~\bibnamefont {Jeffers}}, \bibinfo {author} {\bibfnamefont
  {M.}~\bibnamefont {Padgett}}, \ and\ \bibinfo {author} {\bibfnamefont
  {D.}~\bibnamefont {Faccio}},\ }\href {\doibase
  10.1103/PhysRevLett.117.023601} {\bibfield  {journal} {\bibinfo  {journal}
  {Phys. Rev. Lett.}\ }\textbf {\bibinfo {volume} {117}},\ \bibinfo {pages}
  {023601} (\bibinfo {year} {2016})}\BibitemShut {NoStop}%
\bibitem [{\citenamefont {Chen}\ \emph {et~al.}(2018)\citenamefont {Chen},
  \citenamefont {Ecker}, \citenamefont {Wengerowsky}, \citenamefont {Bulla},
  \citenamefont {Joshi}, \citenamefont {Steinlechner},\ and\ \citenamefont
  {Ursin}}]{chen2018polarization}%
  \BibitemOpen
  \bibfield  {author} {\bibinfo {author} {\bibfnamefont {Y.}~\bibnamefont
  {Chen}}, \bibinfo {author} {\bibfnamefont {S.}~\bibnamefont {Ecker}},
  \bibinfo {author} {\bibfnamefont {S.}~\bibnamefont {Wengerowsky}}, \bibinfo
  {author} {\bibfnamefont {L.}~\bibnamefont {Bulla}}, \bibinfo {author}
  {\bibfnamefont {S.~K.}\ \bibnamefont {Joshi}}, \bibinfo {author}
  {\bibfnamefont {F.}~\bibnamefont {Steinlechner}}, \ and\ \bibinfo {author}
  {\bibfnamefont {R.}~\bibnamefont {Ursin}},\ }\href {\doibase
  10.1103/PhysRevLett.121.200502} {\bibfield  {journal} {\bibinfo  {journal}
  {Phys. Rev. Lett.}\ }\textbf {\bibinfo {volume} {121}},\ \bibinfo {pages}
  {200502} (\bibinfo {year} {2018})}\BibitemShut {NoStop}%
\bibitem [{\citenamefont {Jin}\ and\ \citenamefont
  {Ryosuke}(2018)}]{jin2018Extended}%
  \BibitemOpen
  \bibfield  {author} {\bibinfo {author} {\bibfnamefont {R.~B.}\ \bibnamefont
  {Jin}}\ and\ \bibinfo {author} {\bibfnamefont {S.}~\bibnamefont {Ryosuke}},\
  }\href@noop {} {\bibfield  {journal} {\bibinfo  {journal} {Optica}\ }\textbf
  {\bibinfo {volume} {5}},\ \bibinfo {pages} {93} (\bibinfo {year}
  {2018})}\BibitemShut {NoStop}%
\bibitem [{\citenamefont {Grinin}\ \emph {et~al.}(2020)\citenamefont {Grinin},
  \citenamefont {Matveev}, \citenamefont {Yost}, \citenamefont {Maisenbacher},\
  and\ \citenamefont {Udem}}]{grinin2020Two}%
  \BibitemOpen
  \bibfield  {author} {\bibinfo {author} {\bibfnamefont {A.}~\bibnamefont
  {Grinin}}, \bibinfo {author} {\bibfnamefont {A.}~\bibnamefont {Matveev}},
  \bibinfo {author} {\bibfnamefont {D.~C.}\ \bibnamefont {Yost}}, \bibinfo
  {author} {\bibfnamefont {L.}~\bibnamefont {Maisenbacher}}, \ and\ \bibinfo
  {author} {\bibfnamefont {T.}~\bibnamefont {Udem}},\ }\href@noop {} {\bibfield
   {journal} {\bibinfo  {journal} {Science}\ }\textbf {\bibinfo {volume}
  {370}},\ \bibinfo {pages} {1061} (\bibinfo {year} {2020})}\BibitemShut
  {NoStop}%
\bibitem [{\citenamefont {Ozawa}\ and\ \citenamefont
  {Kobayashi}(2012)}]{ozawa2012chirped}%
  \BibitemOpen
  \bibfield  {author} {\bibinfo {author} {\bibfnamefont {A.}~\bibnamefont
  {Ozawa}}\ and\ \bibinfo {author} {\bibfnamefont {Y.}~\bibnamefont
  {Kobayashi}},\ }\href {\doibase 10.1103/PhysRevA.86.022514} {\bibfield
  {journal} {\bibinfo  {journal} {Phys. Rev. A}\ }\textbf {\bibinfo {volume}
  {86}},\ \bibinfo {pages} {022514} (\bibinfo {year} {2012})}\BibitemShut
  {NoStop}%
\bibitem [{\citenamefont {Marchand}\ \emph {et~al.}(2021)\citenamefont
  {Marchand}, \citenamefont {Riemensberger}, \citenamefont {Skehan},
  \citenamefont {Ho},\ and\ \citenamefont {Kippenberg}}]{marchand2021soliton}%
  \BibitemOpen
  \bibfield  {author} {\bibinfo {author} {\bibfnamefont {P.~J.}\ \bibnamefont
  {Marchand}}, \bibinfo {author} {\bibfnamefont {J.}~\bibnamefont
  {Riemensberger}}, \bibinfo {author} {\bibfnamefont {J.~C.}\ \bibnamefont
  {Skehan}}, \bibinfo {author} {\bibfnamefont {J.~J.}\ \bibnamefont {Ho}}, \
  and\ \bibinfo {author} {\bibfnamefont {T.~J.}\ \bibnamefont {Kippenberg}},\
  }\href@noop {} {\bibfield  {journal} {\bibinfo  {journal} {Nature
  Communications}\ }\textbf {\bibinfo {volume} {12}} (\bibinfo {year}
  {2021})}\BibitemShut {NoStop}%
\bibitem [{\citenamefont {Yepiz-Graciano}\ \emph {et~al.}(2020)\citenamefont
  {Yepiz-Graciano}, \citenamefont {Mart\'{i}nez}, \citenamefont {Lopez-Mago},
  \citenamefont {Cruz-Ramirez},\ and\ \citenamefont
  {U'Ren}}]{pablo2020spectrally}%
  \BibitemOpen
  \bibfield  {author} {\bibinfo {author} {\bibfnamefont {P.}~\bibnamefont
  {Yepiz-Graciano}}, \bibinfo {author} {\bibfnamefont {A.~M.~A.}\ \bibnamefont
  {Mart\'{i}nez}}, \bibinfo {author} {\bibfnamefont {D.}~\bibnamefont
  {Lopez-Mago}}, \bibinfo {author} {\bibfnamefont {H.}~\bibnamefont
  {Cruz-Ramirez}}, \ and\ \bibinfo {author} {\bibfnamefont {A.~B.}\
  \bibnamefont {U'Ren}},\ }\href {\doibase 10.1364/PRJ.388693} {\bibfield
  {journal} {\bibinfo  {journal} {Photon. Res.}\ }\textbf {\bibinfo {volume}
  {8}},\ \bibinfo {pages} {1023} (\bibinfo {year} {2020})}\BibitemShut
  {NoStop}%
\end{thebibliography}%

\end{document}